# PARAMETER TUNING OF TIME-FREQUENCY MASKING ALGORITHMS FOR REVERBERANT ARTIFACT REMOVAL WITHIN THE COCHLEAR IMPLANT STIMULUS

*Lidea K. Shahidi*     *Leslie M. Collins*     *Boyla O. Mainsah*

Department of Electrical and Computer Engineering, Duke University, Durham, NC, USA

## ABSTRACT

Cochlear implant users struggle to understand speech in reverberant environments. To restore speech perception, artifacts dominated by reverberant reflections can be removed from the cochlear implant stimulus. Artifacts can be identified and removed by applying a matrix of gain values, a technique referred to as *time-frequency masking*. Gain values are determined by an oracle algorithm that uses knowledge of the undistorted signal to minimize retention of the signal components dominated by reverberant reflections. In practice, gain values are estimated from the distorted signal, with the oracle algorithm providing the estimation objective. Different oracle techniques exist for determining gain values, and each technique must be parameterized to set the amount of signal retention. This work assesses which oracle masking strategies and parameterizations lead to the best improvements in speech intelligibility for cochlear implant users in reverberant conditions using online speech intelligibility testing of normal-hearing individuals with vocoding.

*Index Terms*— Cochlear implants, Time-frequency mask, Reverberation, Speech enhancement

## 1. INTRODUCTION

Reverberant environments can degrade the intelligibility of speech for both normal-hearing (NH) listeners and hearing-impaired (HI) listeners, such as cochlear implant (CI) users [1]–[3]. Reverberation occurs when sound reflects off of surfaces in an enclosure. When reverberant reflections arrive at a listener at the same time as the signal's direct path from the source, the direct path signal becomes distorted. For speech, energy from reverberant reflections obscures gaps between phonemes and degrades active speech segments, blurring the temporal and spectral cues that facilitate speech understanding. CI users experience great degradations in speech intelligibility in reverberation due to the limited temporal and spectral information provided by the CI after speech processing [4].

To remove signal distortions caused by adverse listening conditions, a matrix of gain values can be applied to a time-frequency (T-F) representation of the distorted signal. This approach to speech enhancement is referred to as *time-frequency masking*. Typically, time-frequency masking is used as a preprocessing step and the time-domain waveform is resynthesized after mask application. Here, we consider using gain values to identify and remove pulses dominated by reverberant distortions directly within CI signal processing, avoiding distortions often imposed by time-domain resynthesis. To assess performance bounds for time-frequency masking strategies, gain values can be determined by an oracle algorithm using a criterion that quantifies the relative amount of the desired and interfering signal, in this case the direct path and the reverberant reflections, respectively. There are two major types of T-F masks: the *ideal binary mask* (IBM) and the *ideal ratio mask* (IRM) [5], [6]. The IBM contains only binary gain values, 0 or 1, leading to deletion or retention of T-F units of the distorted signal, respectively. The IRM contains gain values that assume continuous values between 0 and 1.

T-F masking is an effective speech enhancement technique for noisy speech [7]–[9] and recent efforts have extended T-F masking to improve speech intelligibility in reverberant conditions for NH and HI listeners [2], [10]–[15]. To date, however, it is unclear which masking strategy would most benefit reverberant speech intelligibility for CI users when masks are applied at the pulse level. When comparing IRM- and IBM-mitigated signals in noise, the IRM resulted in better speech intelligibility outcomes for NH listeners with vocoding but no significant difference in speech intelligibility outcomes for CI users [16]. Whether similar results are applicable to reverberant speech remains undetermined. The effectiveness of T-F masking is also mediated by the parameterization of each masking strategy. The IBM and IRM contain parameters that impact the amount of signal retention and therefore the efficacy of the resulting mask. Typically, studies employing T-F masking strategies select parameterizations without specifying the selection criteria [11], [12], [14], [15]. Some studies have explored a range of mask parameterizations for the IBM in reverberation, either with NH listeners [10] or CI users [2], indicating parameterizations that led to better speech intelligibility outcomes. Although these studies established a general framework for evaluating mask parameterizations using subjective testing, they considered only one mask type, employ T-F masking as a pre-processing step, and utilized somewhat unrealistic reverberant conditions which did not incorporate the reflective characteristics of surfaces found in real rooms.

T-F masking strategies have the potential to benefit CI users in reverberant conditions. Using realistic reverberant conditions, this work will compare the efficacy of different T-F masking models and various model parameterizations when masks are applied within CI processing. Online intelligibility of reverberant speech mitigated by masks spanning a range of realistic parameterizations will be tested in NH-listeners with vocoded signals.

## 2. TIME-FREQUENCY MASKING IN REVERBERANT CONDITIONS

Reverberation can be modelled as the convolution of a room impulse response (RIR) with the anechoic signal. The RIR is the transfer function between the anechoic signal and the reverberant signal. The reverberant signal $y(t)$ is defined as

$$y(t) = h(t) * s(t), \quad (1)$$

where $s(t)$ is the anechoic signal; and $h(t)$ is the RIR between the source and receiver. Figure 1 illustrates a recorded RIR composed of a series of impulses. The initial impulse of the RIR indicates the *direct path* of the reverberant signal, or a delayed and attenuated copy of the source signal that is unperturbed by reverberant reflections [17]. The direct path signal $d(t)$ can be estimated accordingly:

$$d(t) = h_d(t) * s(t), \quad (2)$$

where $h_d(t)$ is the direct path component of the RIR. Since the direct path signal is undistorted by reverberant reflections, it is likely to be as intelligible as the anechoic signal. The impulses following the initial impulse of the RIR, as shown in Figure 1, indicate reverberant reflections.

To quantify the amount of reverberant distortion, we use the speech-to-reverberant ratio (SRR) over T-F units of the reverberant signal, where SRR is the local ratio of the energy of the direct path signal to the energy of the residual reverberant reflections [17]. A short-time Fourier transform can be used to obtain T-F representations of the reverberant and direct path signals, which we will denote as $D(t,f)$ and $Y(t,f)$, respectively, where $t$ and $f$ represent time and frequency, respectively. The SRR of a T-F unit (expressed in dB), can be estimated accordingly:

$$SRR(t,f) = 10 \log_{10} \left( \frac{|D(t,f)|^2}{|Y(t,f) - D(t,f)|^2} \right). \quad (3)$$

SRRs can take any real number value, although a preliminary analysis of SRR values over a range of reverberant conditions and speakers suggest that they typically range from -75 to 75 dB. The SRR assumes negative values when the energy of the reverberant reflections in a T-F unit is greater than the energy of the direct path signal and assumes positive values when the energy in the T-F unit is dominated by the direct path signal.

To retain signal components predominantly containing the direct path signal, the SRR can be thresholded, resulting in the *ideal binary mask* (IBM) [2], [10]:

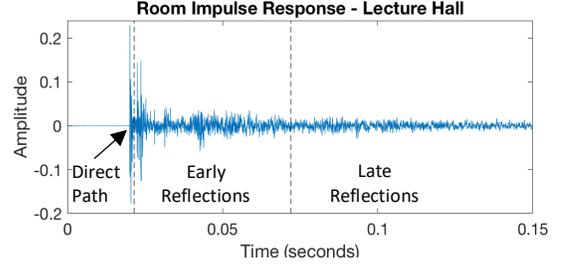

**Fig. 1.** An example room impulse response (RIR) segmented to illustrate the impulse corresponding to the direct path signal, early and late reverberant reflections. The RIR was truncated from 1.4 seconds to 0.15 seconds to better visualize the RIR components.

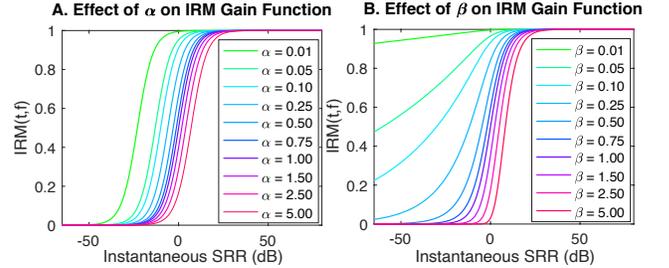

**Fig. 2.** Gain function curves mapping SRR values to IRM values (A) with varying $\alpha$ at $\beta = 1$; and (B) with varying $\beta$ at $\alpha = 1$.

$$IBM(t,f) = \begin{cases} 1 & if\ SRR(t,f) > \tau_l \\ 0 & otherwise \end{cases} \quad (4)$$

where $\tau_l$ is the local threshold in dB. The local threshold is the target threshold, $\tau$, adjusted by the effective SNR ($\tau_l = \tau + eSNR$) [10]. The effective SNR (eSNR) is the energy in the time domain of the direct path signal over the reverberant reflections (in dB) estimated accordingly [10].

$$eSNR = 10 \log_{10} \left( \frac{\sum_t d(t)^2}{\sum_t (y(t) - d(t))^2} \right) \quad (5)$$

The effective SNR of a speech signal can vary even within speech material spoken by the single speaker in the same reverberant conditions. Inclusion of eSNR allows the choice of $\tau$ to be independent of the relative energy in the speech material considered.

As an alternative to thresholding the SRRs to generate a binary mask, a sigmoidal function can be used to map SRR values to continuous gain values between 0 and 1 to generate an *ideal ratio mask* (IRM) accordingly [11], [18]:

$$IRM(t,f) = \left( \frac{lSRR(t,f)}{lSRR(t,f) + \alpha} \right)^{\beta}, \quad (6)$$

where $lSRR(t,f)$ is the SRR on a linear scale (i.e., the operand of the logarithm function in (3)); and $\alpha$ and $\beta$ control the shift and slope, respectively, of the gain function, as illustrated in Figure 2. In most studies utilizing the IRM for mitigating reverberation, $\alpha$ and $\beta$ are typically set to 1 and 0.5, respectively [11].

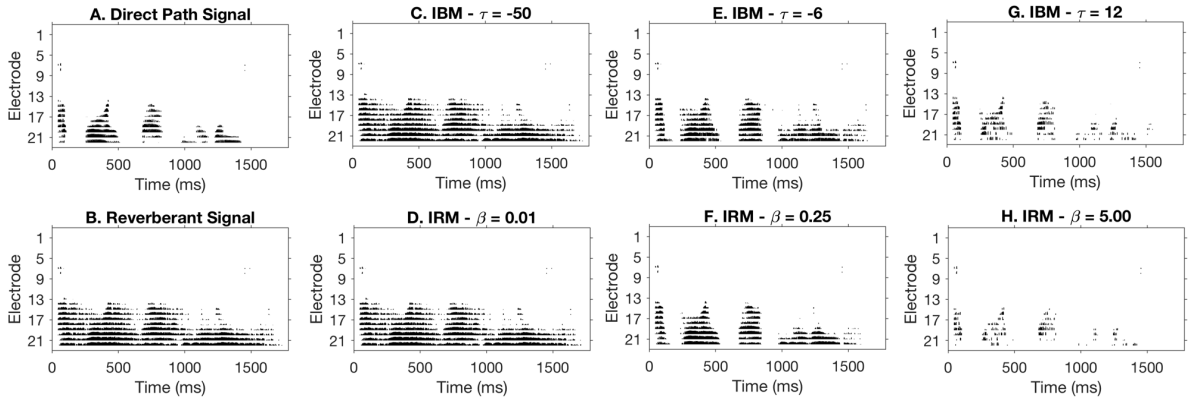

**Fig. 3.** Electrodograms of the speech token "A boy fell from the window," showing the CI stimulation pattern of (A) the direct path signal; (B) the reverberant signal; and mitigated reverberant signals obtained by applying (C, E, G) ideal binary masks (IBMs) and ideal ratio masks (IRMs) with increasing parameter values, $\tau$ and $\beta$, respectively, to the reverberant signal. The mask parameter controls the sparsity of the mask, resulting in under-attenuation (C and D) to over-attenuation (G and H) of the reverberant signal as the parameter value increases.

The choice of mask parameter values can impact the effectiveness of reverberation mitigation as they control the sparsity of the T-F mask, and consequently the amount of retention of the reverberant CI stimulus. Figure 3 illustrates the reverberant CI stimuli before and after the application of T-F masks with different parameterizations. Under-attenuation of reverberant signals due to sparse masks generated from low parameter values results in mitigated signals that contain large amounts of reverberant artifacts between phonemes (Figure 3 C and D) potentially degrading the spectro-temporal cues that are beneficial for speech intelligibility. Over-attenuation of reverberant signals due to sparse masks generated with high parameter values results in sparse signals where important speech information may have been removed (Figure 3 G and H) potentially degrading the signal intelligibility.

It remains an open question whether alternative parameterizations of $\tau$ for the IBM and $\beta$ for the IRM would benefit speech intelligibility for CI users in reverberant conditions. Towards answering this question, this study will explore the speech intelligibility outcomes of different mask parameterizations for NH listeners with vocoding.

## 3. METHODS

This experiment compared the effectiveness of T-F masking strategies applied within CI processing over a range of parameterizations within each mask type. To mitigate reverberation, a binary or ratio mask was generated for each parameter value and applied to T-F units of a reverberant CI stimulus pattern, and the resulting speech intelligibility of the mitigated reverberant signal assessed. For comparison, the speech intelligibilities of the unmitigated reverberant signal and the direct-path reverberant signal (2) were also tested.

Speech material consisted of sentences from the HINT speech corpus spoken by a single male speaker [19]. The HINT corpus consists of 25 phonemically balanced lists. Each list contains 10 sentences, with 6 to 7 syllables in length. Prior to processing, the speech signals were first downsampled from 22160 Hz to 16000 Hz. Reverberant speech material was created by convolving anechoic speech material with a RIR function (1) from the Aachen Impulse Response (AIR) database [20]. The RIR used in this experiment had a reverberation time ($RT_{60}$) of 0.8 seconds and was recorded in a lecture hall with a room length of 10.8m, width of 10.9m and height of 3.15m. The source-to-microphone distance was 7.1 m, which is well outside the critical distance of the room, approximated as 1.2 meters [17]. This reverberant condition was chosen as it provides a realistic scenario in which a CI user would likely have difficulty understanding speech [2], [3]. To create the direct-path signal, the RIR was truncated to remove impulses occurring 5 milliseconds after the initial impulse and convolved with the anechoic speech material. The root-mean-square (RMS) level of all reverberant speech material was normalized to the highest RMS that avoided clipping.

Reverberant signals were passed through a simulation of the Advanced Combination Encoder (ACE) processing [21] (implemented using the Nucleus Matlab Toolbox [22]) resulting in reverberant stimulus patterns. T-F masks were computed for T-F representations of the reverberant signal obtained after envelope extraction and applied to the reverberant CI stimulus patterns prior to channel selection. IBMs were created for each $\tau$ value (4) while IRMs were created for each $\beta$ value with α set to 1 (6). We set $\alpha = 1$ based on the intuition that the magnitudes of the direct path signal and the reverberant reflections should be on the same scale [11]. As described in Section 2, the thresholds ($\tau$) applied when creating the IBMs were adjusted by the eSNR (5) to account for variations in eSNR across different sentences within the same threshold condition. Vocoded waveform stimuli of the mitigated reverberant stimulus pattern were generated by using the resulting CI pulse amplitudes as the amplitudes of sinusoidal carriers, with the frequencies of the sinusoidal carriers corresponding to the frequencies of electrode channels in the default ACE map.

Twenty native speakers of American English were recruited to participate in this study, which was approved by Duke University Institutional Review Board. The subject ages ranged from 19 to 35 years. All subjects had self-reported normal hearing and were paid for their participation.

During the test, subjects were seated at a computer in a soundproof booth. Vocoded speech stimuli were presented to the listener diotically through Sony MDR7506 headphones at a sound-pressure level (SPL) of 65 dB. Each sentence was presented once, and subjects were instructed to type the words they were able to hear. A study administrator was seated in the booth with the subject during the study.

Each session began with an adaptive training task to familiarize subjects with listening to vocoded speech. Subjects were presented with feedback and training ended when their performance plateaued over five sentences or when thirty sentences were presented. After training, the direct-path condition and the unmitigated condition were presented. Ratio mask conditions and binary mask conditions were then presented in blocks, with $\tau$ and $\beta$ conditions randomized within each block. The presentation order of ratio mask and binary mask blocks was counterbalanced across subjects. Each task used a randomly selected HINT list, sampled without replacement. Each session lasted 45-60 minutes, including breaks.

## 4. RESULTS

Figure 4 shows the speech intelligibility results, measured as the percent of correct phonemes, for IBM and IRM conditions across $\tau$ and $\beta$ values, respectively. The direct-path condition resulted in recognition accuracy of 91.2% ± 6.6%, which is similar to previously reported speech intelligibility performance with vocoded speech in NH users under anechoic conditions [23]. The unmitigated condition resulted in an accuracy of 11.2% ± 9.1%, confirming the difficulty of the reverberant condition. For the mitigated reverberant conditions, there was a concave relationship between speech intelligibility and mask parameter value for both T-F masking strategies, suggesting that speech intelligibility reflects a balance between retention of active speech and deletion of detrimental reverberant signal components.

A repeated measures ANOVA test was performed to assess the effectiveness of mask parameterization on speech intelligibility, followed by *post hoc* Tukey's HSD (honestly significant difference) test for pairwise comparisons. Subject mean scores were first transformed using the rationalized arcsine transform [24] prior to the ANOVA analysis. A significant effect of mask parameter value for both the IRM [$F(10,190) = 50.1$, $p < 0.0001$] and the IBM [$F(10,190) = 50.02$, $p < 0.0001$] was observed, with large partial $\eta^2$ effect sizes for group mean differences of 0.7250 and 0.7247, respectively. The pairwise comparisons revealed that the unmitigated condition was significantly different ($p < 0.05$) from mitigated reverberation conditions for a subset of parameter values for both masks (see Figure 4). Averaging across subjects, the maximum performance was achieved at $\tau = -6$ dB for the IBM and $\beta = 0.25$ for the IRM, with speech intelligibility raised by 40.2% and 31.1%, respectively, from the unmitigated condition. These results demonstrate the sensitivity of T-F masking approaches to the choice of parameter settings.

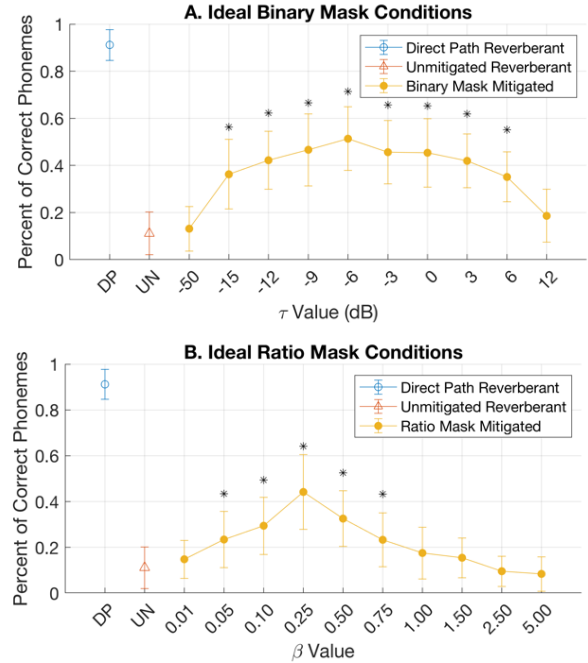

**Fig. 4.** Percent of phonemes correctly identified in (A) IBM and (B) IRM conditions as a function of $\tau$ or $\beta$, respectively. Each marker indicates the average score over twenty listeners, with error bars indicating the standard deviation. Performance given the direct path of the reverberant signal (DP) and the unmitigated reverberant signal (UN) are given to the left in each plot. Asterisks above mask conditions indicates significant pairwise differences from the unmitigated condition.

## 5. DISCUSSION

This study investigated the potential of applying different time-frequency masking strategies within CI processing and the impact of mask parameterization on the resulting speech intelligibility in reverberant conditions. By testing over parameter values which ranged from under- to over-attenuation of the reverberant signal, we demonstrated that parameters can be selected which lead to a level of attenuation that maximizes speech intelligibility. We found a balance between signal retention and removal using a threshold value of −6 dB for binary masking conditions, in agreement with studies of binary mask parameterization in reverberation recommending threshold values of −6 dB for NH listeners [10] and −5 dB for CI users [2]. To our knowledge, this analysis of ratio masking slope parameters is the only such exploration in reverberant conditions. In general, IBM-mitigated signals resulted in larger improvements in speech intelligibility than IRM-mitigated signals. Anecdotal evidence from an informal survey of half of the subjects revealed an overwhelming preference (9 out of 10 subjects) for the binary mask conditions in terms of speech quality. These results suggest that for mitigating reverberant signals with a time and frequency resolution similar to that of CI stimuli, the binary mask may lead to better improvements in speech intelligibility than the ratio mask.

Additional efforts are needed to create a T-F masking mitigation strategy that can be implemented in real world conditions. A real world T-F masking strategy would require blind estimation of the SRR from the reverberant signal while this study assumed knowledge of the RIR to estimate the SRR. This work supplies an informed parameterization objective for a potential machine learning paradigm that aims to estimate T-F masking strategies in real-world scenarios. A machine learning algorithm could be trained to estimate IBMs using only knowledge of the reverberant signal, with ground truth binary masks based on IBMs generated at a threshold of -6 dB.

## 6. ACKNOWLEDGEMENT

The authors would like to thank the subjects who participated in this experiment. This work was funded by the National Institute of Health, via a grant administered by the National Institute on Deafness and Other Communication Disorders (#R01-DC014290).